\documentclass[a4paper,fleqn,usenatbib]{mnras}

\usepackage{newtxtext,newtxmath}

\usepackage[T1]{fontenc}
\usepackage{ae,aecompl}

\usepackage{graphicx}	

\usepackage{amsmath}	
\usepackage{amssymb}	

\title[New results on Cycles 22 -- 25 from BiSON]{Sub-surface structural changes associated with successive 11-yr solar activity cycles have been progressively more confined near the surface: new helioseismic results on Cycles 22 -- 25 from BiSON}

\author[W. J. Chaplin et~al.]{William J. Chaplin$^{1}$\thanks{E-mail: w.j.chaplin@bham.ac.uk}, Sarbani Basu${^2}$\thanks{E-mail: sarbani.basu@yale.edu}, Rachel Howe$^1$, Yvonne Elsworth$^1$,
\newauthor Steven J. Hale$^1$, Eleanor Murray$^1$\\
$^{1}$School of Physics and Astronomy, University of Birmingham, Birmingham, B15 2TT, United Kingdom\\
$^{2}$Department of Astronomy, Yale University, PO Box 208101, New Haven, CT, 065208101, USA}

\date{Accepted XXX. Received YYY; in original form ZZZ}

\pubyear{2026}

\begin{document}
\label{firstpage}
\pagerange{\pageref{firstpage}--\pageref{lastpage}}
\maketitle

\begin{abstract}

We use Sun-as-a-star helioseismology data, collected by the Birmingham Solar-Oscillations Network (BiSON), to examine the relationship between the solar-cycle-induced frequency shifts of whole-Sun, low-angular degree solar p modes and well-known proxies of global solar activity. Changes in behaviour between the low-frequency modes and proxies, which in a previous study we found had occurred on the declining phase of Cycle~23, appear to have persisted into Cycle~25. More striking is a significant change in the relationship for higher-frequency modes, which the new Cycle~25 data now reveal. The observed mean frequency shifts in Cycle~25 are much stronger than one would expect for these modes based on the relationship between the frequencies and proxies seen in previous cycles, in particular Cycle~22. In sum, Cycle 25 is as strong as Cycles~22 and 23 when observed in this higher-frequency seismic band, in marked contrast to the relative sizes of the cycles seen in the global activity proxies, where Cycle~25 is noticeably weaker. When considered alongside a systematic reduction of the sensitivity of the mid-frequency modes to activity over the past three cycles, these results suggest that sub-surface structural changes associated with successive 11-year cycles are becoming ever more progressively confined just beneath the solar surface. 

\end{abstract}




\section{Introduction}
\label{sec:intro}

The last few solar cycles have seen significant changes in overall levels of activity and differences in the evolution of magnetic fields at different solar latitudes (e.g., see \citealt{Hathaway2015,Norton2023}). Cycle 24 was significantly weaker in well-known proxies of global solar activity than previous cycles, and marked a departure from the preceding so-called modern maximum epoch (\citealt{Usoskin2017}). While the current Cycle 25 has peaked at higher activity, it did not return to pre-Cycle-24 levels.

The frequencies of solar p modes respond to changing levels of magnetic activity, and whole-Sun, low angular-degree (low-$l$) modes provide a truly global seismic diagnostic of solar cycles (e.g., see recent examples in \citealt{Garcia2025, Howe2025}), and the internal structural changes associated with them. In \citet{Basumin2012}, we examined the relationship between the solar-cycle-induced frequency shifts of these low-$l$ p modes and well-known proxies of global solar activity, over an epoch that spanned a good fraction of Cycle 21 through to the very early part of Cycle 24. The helioseismic data came from Sun-as-a-star Doppler velocity observations made by the long-standing Birmingham Solar-Oscillations Network (BiSON; \citealt{Chaplin1996, Hale2016}). We examined the p-mode frequencies by averaging them over different frequency bands, and found that there was a marked change in the mode-proxy relationship for modes of frequency less than $2400\,\rm \mu Hz$. This change occurred during the declining phase of Cycle 23, and from it, we inferred changes in solar structure confined within a sub-surface layer shallower than $\approx 3000\,\rm km$. The changes must have been in deeper-lying layers in the preceding Cycle 22. In \citet{Howe2017}, we included several more years of BiSON data, and found that the above-mentioned changed relationship appeared to have persisted through the rising phase of Cycle 24.

The BiSON dataset now covers the rest of Cycle 24, and the full rising phase of Cycle 25. Our first objective in this paper is therefore to test whether the changed behaviour of the low-frequency modes has continued into the new cycle. Our second objective is to test whether indications of possible changes in behaviour at higher frequencies in Cycles 23 and 24, which had been flagged by \citet{Howe2017}, are real and have grown to significant levels.

The layout of the rest of the paper is as follows. We introduce the BiSON and activity proxy data in Section~\ref{sec:data}, and present overlapping 1-yr average frequency shifts in different frequency bands, proxy averages for the same epochs, and results from linear regressions of the average shifts on the average proxies for each cycle. We explore the changing relationship between the frequency shifts and proxies in Section~\ref{sec:res}. We finish the paper by summarising our main findings in Section~\ref{sec:conc}.

\section{Data and Analysis}
\label{sec:data}

We have used Doppler velocity data collected by the BiSON telescopes between 1987 and 2025. This period begins with the rising phase of Cycle 22, and ends at the maximum of Cycle 25. Data from each site were calibrated and combined into a single, concatenated timeseries, using the procedures described by 
\citet{Davies2014}. We then analysed the concatenated set in overlapping segments of length 1\,yr, with consecutive segments offset in time by 3\,months. Mode frequencies $\nu_{nl}$ of degree $l$ and overtone number $n$ were extracted from the frequency power spectrum of each 1-yr segment using the Bayesian fitting algorithm described by \citet{Howe2023}.

\begin{figure*}
\centering
\includegraphics[width=0.65\textwidth]{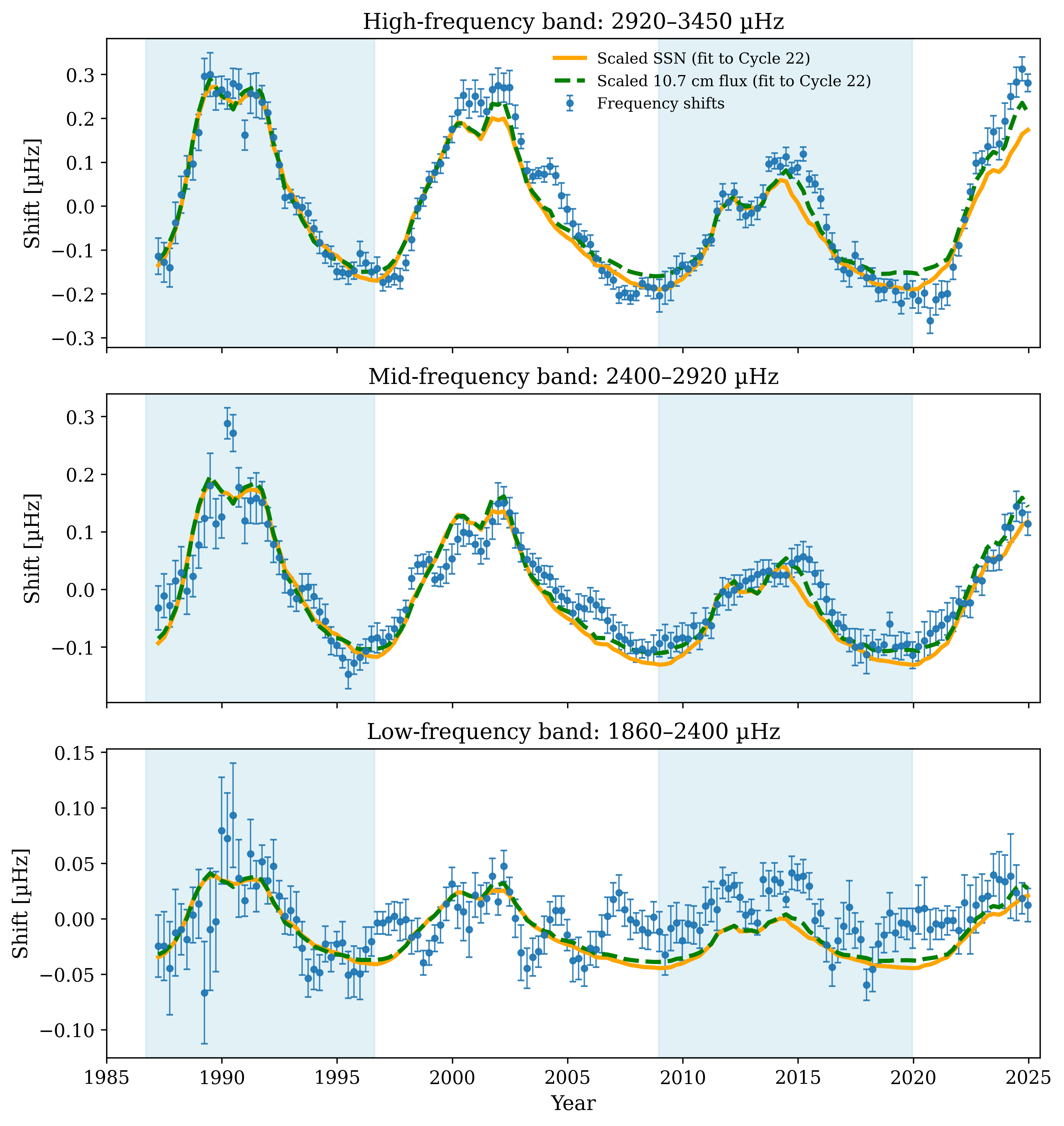}
        \caption{Points with error bars: mean frequency shifts for each of the overlapping 1-yr segments in three different frequency bands, covering low (bottom panel), mid (middle panel) and high (top panel) ranges. Solid orange and dashed black lines: 1-yr averages of the 10.7-cm radio flux and SSN, respectively, scaled for the plots in each panel by a linear fit to the appropriate band frequency shifts in Cycle 22.}
	  \label{fig:rawshifts}
\end{figure*}

Following \citet{Basumin2012} and \cite{Howe2017}, we calculated the mean frequency shifts for each of the overlapping 1-yr segments in three different frequency bands, covering low ($1860 \le \nu_{nl} \le 2400\,\rm \mu Hz$), mid ($2400 \le \nu_{nl} \le 2920\,\rm \mu Hz$) and high ($2920 \le \nu_{nl} \le 3450\,\rm \mu Hz$) ranges. Each mean shift was computed with respect to the temporal mean of the frequencies of all modes in the targeted frequency range (e.g., see \citealt{Howe2015}). The mean shifts are shown in Fig.~\ref{fig:rawshifts} (points with error bars). Also plotted are two commonly used proxies of global solar activity, the 10.7-cm radio flux (\citealt{Tapping2013}) and the Sunspot Number (SSN), here from the revised Brussels–Locarno catalogue (\citealt{Clette2016, Clette2023}). Their daily estimates were averaged over the same epochs as the 1-yr BiSON segments, and scaled for the plot by linear fits (using independent data only) of the mean shifts in each band to the averaged proxies in Cycle 22. Anchoring the scaling to that observed in Cycle 22 provided a suitable reference against which to visually show how the relationship between the mean frequency shifts and the activity proxies has changed from one cycle to another, and is the approach we adopted for visual comparison in our previous papers. Taking the SSN as an example, the linear model took the form $\langle \delta \nu_l(t) \rangle = a_{\rm SSN} \langle {\rm SSN}(t) \rangle + c_{\rm SSN}$, where $\langle \delta \nu_l(t) \rangle$ are the mean shifts at degree $l$, $\langle {\rm SSN}(t) \rangle$ are the mean SSN data, and $a_{\rm SSN}$ and $c_{\rm SSN}$ are, respectively, the linear gradient and offset. The scaled proxy averages $\langle {\rm SSN}'(t) \rangle$ for the plots were then calculated from the best-fitting gradient and offset, i.e., from $\langle {\rm SSN}'(t) \rangle = a_{\rm SSN} \langle {\rm SSN}(t) \rangle + c_{\rm SSN}$. In summary, if the relationship between the frequency shifts and proxies is different in later cycles compared to Cycle 22, we will see the frequency shifts and proxies depart from one another on the plots.

In order to quantify those changes, we performed linear fits of the mean shifts to the average proxies for each of cycles 23, 24 and 25 individually, in addition to the above-mentioned fit to Cycle 22. Table~\ref{tab:grads} shows the fitted linear gradients, $a_{\rm F10.7}$ and $a_{\rm SSN}$ (in units of nHz per proxy) and their corresponding $p$ values, by frequency band; and the gradient differences $\Delta a$ with respect to the Cycle 22 gradients, normalised by the uncertainties on the differences $\sigma_{\Delta a}$ given by combining the gradient uncertainties in quadrature.

\begin{figure*}
\centering
\includegraphics[width=0.65\textwidth]{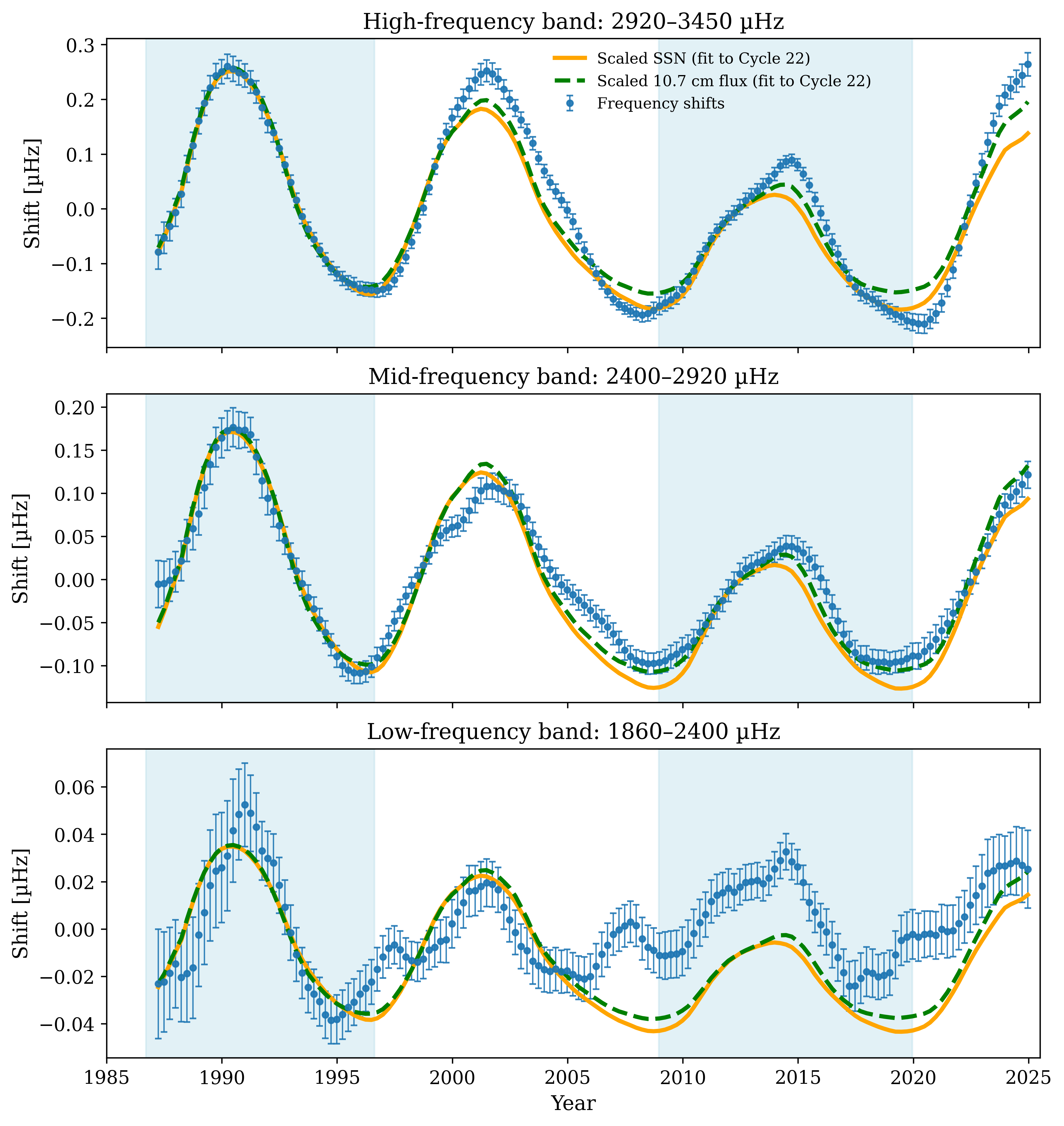}
        \caption{Mean frequency shifts and activity proxies for each of the overlapping 1-yr segments, as per Fig.~\ref{fig:rawshifts} but now after applying a 9-point (2.25-yr) boxcar filter to smooth over the shorter-period, smaller amplitude quasi-biennial signal.}
	  \label{fig:smshifts}
\end{figure*}

The mean frequency shifts (and proxies) also contain a shorter-period (approx. 2\,yr), smaller-amplitude signal called the  ``quasi biennial'' variation (e.g., see \citealt{Fletcher2010, Broomhall2012, Simoniello2012, Jain2023}). In order to accentuate the longer-period signatures of the 11-yr cycle, we also followed \citet{Basumin2012} and \cite{Howe2017} and smoothed over the quasi-biennial variation by applying a 9-point (2.25-yr) boxcar filter to the mean shifts and mean proxies. Fig.~\ref{fig:smshifts} plots the resulting smoothed data. (Note that fits to these smoothed data -- where we used a Monte-Carlo bootstrapping approach to estimate associated uncertainties -- returned similar results to the fits on the unsmoothed, independent shifts presented in Table~\ref{tab:grads}.)


\begin{table*}
\centering
\caption{Results of linear regression fits of the independent frequency shifts to the activity proxies in each cycle. Gradients $a$ are given in nHz per proxy unit, along with their corresponding $p$ values. The quantity $\Delta a/\sigma_\Delta$ gives the difference of each cycle's gradient from the Cycle\,22 value in units of the combined uncertainty. Results are shown for fits against the 10.7-cm radio flux (F10.7) and Sunspot Number (SSN).}
\label{tab:grads}
\begin{tabular}{ccccccc}
\hline
Cycle &$a_{\rm F10.7}$ & $p(a_{\rm F10.7})$ & $\Delta a/\sigma_{\Delta a}$ (F10.7) &$a_{\rm SSN}$   & $p(a_{\rm SSN})$ & $\Delta a/\sigma_{\Delta a}$ (SSN) \\
\hline
\multicolumn{7}{c}{\textsl{High-frequency band ($2920 \le \nu_{nl} \le 3450\,\rm \mu Hz$)}} \\
\noalign{\smallskip}
22 & $3.31 \pm 0.21$ & $2.61\times 10^{-7}$ & -- & $2.36 \pm 0.17$ & $5.51\times 10^{-7}$ & -- \\
23 & $3.85 \pm 0.36$ & $8.19\times 10^{-7}$ & $+1.30$ & $2.56 \pm 0.27$ & $2.65\times 10^{-6}$ & $+0.65$ \\
24 & $4.16 \pm 0.39$ & $1.75\times 10^{-6}$ & $+1.94$ & $2.72 \pm 0.42$ & $1.24\times 10^{-4}$ & $+0.79$ \\
25 & $4.40 \pm 0.33$ & $8.44\times 10^{-4}$ & $+2.77$ & $3.39 \pm 0.33$ & $1.92\times 10^{-3}$ & $+2.80$ \\
\noalign{\smallskip}\hline\noalign{\smallskip}
\multicolumn{7}{c}{\textsl{Mid-frequency band ($2400 \le \nu_{nl} \le 2920\,\rm \mu Hz$)}} \\
\noalign{\smallskip}
22 & $2.49 \pm 0.37$ & $1.32\times 10^{-4}$ & -- & $1.76 \pm 0.25$ & $1.15\times 10^{-4}$ & -- \\
23 & $1.61 \pm 0.25$ & $6.97\times 10^{-5}$ & $-2.00$ & $1.09 \pm 0.17$ & $7.19\times 10^{-5}$ & $-2.20$ \\
24 & $1.91 \pm 0.33$ & $2.27\times 10^{-4}$ & $-1.19$ & $1.23 \pm 0.24$ & $5.90\times 10^{-4}$ & $-1.52$ \\
25 & $1.77 \pm 0.27$ & $6.60\times 10^{-3}$ & $-1.60$ & $1.35 \pm 0.22$ & $8.05\times 10^{-3}$ & $-1.23$ \\
\noalign{\smallskip}\hline\noalign{\smallskip}
\multicolumn{7}{c}{\textsl{Low-frequency band ($1860 \le \nu_{nl} \le 2400\,\rm \mu Hz$)}} \\
\noalign{\smallskip}
22 & $0.63 \pm 0.26$ & $4.37\times 10^{-2}$ & -- & $0.44 \pm 0.19$ & $4.63\times 10^{-2}$ & -- \\
23 & $0.26 \pm 0.19$ & $2.13\times 10^{-1}$ & $-1.13$ & $0.15 \pm 0.14$ & $3.01\times 10^{-1}$ & $-1.22$ \\
24 & $0.81 \pm 0.24$ & $8.84\times 10^{-3}$ & $+0.50$ & $0.52 \pm 0.17$ & $1.44\times 10^{-2}$ & $+0.29$ \\
25 & $0.30 \pm 0.21$ & $2.43\times 10^{-1}$ & $-0.97$ & $0.23 \pm 0.16$ & $2.53\times 10^{-1}$ & $-0.86$ \\
\noalign{\smallskip}
\hline
\end{tabular}
\end{table*}


\section{Results}
\label{sec:res}

We first remark on the behaviour in the low-frequency band. A change in the relationship between the low-frequency shifts and the activity proxies -- identified by \citet{Basumin2012} as having set in during Cycle 23, and observed by \citet{Howe2017} as having remained on the rising phase of Cycle 24 -- still appears to be present into Cycle 25. Figs.~\ref{fig:rawshifts} and~\ref{fig:smshifts} show a continued departure of the low-frequency shifts from the proxies, beginning circa. 2005, when the proxies are scaled according to the shift-proxy relationship in Cycle 22. This departure is seen largely in the form of an offset, i.e., from 2005 to 2010 the frequencies did not decrease like the scaled proxies, and then at later epochs continued to change from a higher base-level. Note that this behaviour is still seen if we scale the proxies in a different way, e.g., by first subtracting an average proxy computed across the entire Cycle 22 to 25 period, and then scaling the proxy residuals by the best-fitting gradient from the linear fit in Cycle 22 only. The sensitivity gradients $a_{\rm F10.7}$ and $a_{\rm SSN}$ and $p$-values (Table~\ref{tab:grads}) were very weak in Cycles 23 and 25, compared to Cycle 22, though they did increase in Cycle 24 (that increase is not as large if only the rising phase of Cycle 24 is considered).

It is also now apparent that in the mid-frequency band, the sensitivity gradients have been consistently lower in all of Cycles 23, 24 and 25 compared to Cycle 22. The probability of obtaining three negative deficits in $\Delta a/\sigma_{\Delta a}$ of this size or larger by chance (one-sided test) is 0.015\,\% for $a_{\rm F10.7}$, and 0.010\,\% for $a_{\rm SSN}$; whilst that of obtaining three absolute differences in $\Delta a/\sigma_{\Delta a}$ of this size or larger by chance (two-sided test) is 0.12\,\% for $a_{\rm F10.7}$, and 0.08\,\% for $a_{\rm SSN}$. As we shall go on to discuss below, this result has potentially interesting implications for the localisation of the near-surface structural changes driving the shifts, in particular when we consider it alongside what is arguably the most striking feature of the figures: the marked change in behaviour we now see in the high-frequency band on the rising phase of Cycle 25. 

The observed high-frequency shifts are much stronger in Cycle 25 than one would expect based on the relationship between the frequencies and proxies seen in Cycle 22. Note how in Cycle 25, the frequency shifts show a larger minimum-to-maximum swing than, and peak well above, the scaled proxies. In summary, Cycle 25 appears as strong as Cycles 22 and 23 in this seismic frequency band. This is in marked contrast to the relative sizes of the cycles seen in the activity proxies, e.g., when smoothed over 1\,yr, the peak of the SSN cycle is $\approx$25\,\% weaker in Cycle 25 than in Cycle 22.

\citet{Howe2017} had already remarked that the sensitivity of the mode frequencies to the activity proxies was slightly higher in Cycles 23 and 24 than in Cycle 22. That behaviour is evident here; however, it has now become much more pronounced in Cycle 25. The sensitivity of the high-frequency modes to the activity proxies is $\approx 3\sigma$ higher in Cycle 25 than it was in Cycle 22. Fig.~\ref{fig:smshifts} in particular suggests that the change in the high-frequency relationship may have set in as early as the maximum of Cycle 23. Irrespective, the difference in Cycle 25 is reasonably clear, and fairly pronounced.

By considering the sensitivity of the modes to structural changes at different solar radii, we may comment on what the above results mean for how these changes have been localised in radius. The high-frequency band modes are much more sensitive to changes confined within $\simeq 1000\,\rm km$ of the surface than are their mid- and low-frequency counterparts; while the low-frequency modes have reduced sensitivity to changes shallower than $\simeq 3000\,\rm km$ (e.g., see Fig.~6 of \citealt{Basumin2012}). Based on the low-frequency behaviour, \citet{Basumin2012} and then \citet{Howe2017} had concluded that structural changes in Cycle 23 were localised mainly within $3000\,\rm km$ of the surface, and that changes in Cycle 22 were deeper-lying in nature. The increased sensitivity of the high-frequency modes to activity that we see in Cycle 25, coupled to the systematic reduction over the past three cycles of the sensitivity of the mid-frequency modes, therefore suggests that sub-surface structural changes driving the shifts have become even more confined close to the surface since Cycle 22. 

\section{Conclusions}
\label{sec:conc}

We have used BiSON Sun-as-a-star helioseismology data to examine the relationship between the solar-cycle-induced frequency shifts of whole-Sun, low-$l$ solar p modes and two well-known proxies of global solar activity, the 10.7-cm radio flux and Sunspot Number. We find that a marked change in behaviour between the low-frequency modes and proxies, which in \citet{Basumin2012} we discovered had occurred on the declining phase of Cycle 23, appears to have persisted into Cycle 25. It is interesting to speculate whether the collection of further data over what remains of Cycle 25 and into the upcoming Cycle 26 might reveal a longer period associated with this behaviour, e.g., one associated with the 22-year Hale cycle on which the Sun reverts to having the same magnetic polarity.

More striking is a significant change we have now uncovered in the relationship for higher-frequency modes, which the new Cycle 25 data reveal. The observed mean high-frequency shifts are found to be much stronger in Cycle 25 than one would expect based on the relationship between the frequencies and proxies seen in previous cycles, in particular Cycle 22. When observed in the high-frequency p modes, Cycle 25 is therefore as strong as Cycles 22 and 23. This is in marked contrast to the relative sizes of the cycles seen in the global activity proxies, where Cycle 25 is noticeably weaker. When considered alongside a systematic reduction of the sensitivity of the mid-frequency modes to activity over the past three cycles, these results suggest that sub-surface structural changes associated with successive 11-year cycles are becoming ever more progressively confined close to the surface. 

It is perhaps worth remarking that other recent studies in the literature have presented evidence for several-decade-long changes in near-surface phenomena, spanning several cycles, e.g., a systematic decrease across Cycles 21 -- 24 in the number of sunspots observed in spot groups (\citealt{Pevtsov2025}), following previous claims of a systematic decline in the average sunspot field strength (\citealt{Nagovitsyn2012}) possibly due to a change in the balance of large and small spots. We note that our results cannot be explained by changes in the strength of the sub-surface field alone: whilst this would alter the absolute sizes of the observed shifts, it would not change the sensitivity of the modes to the activity. A change in the radial confinement of the field is instead needed.

\section*{Acknowledgements}

We would like to thank all those who are, or have been, associated with the Birmingham Solar-Oscillations Network (BiSON), in particular P. Pallé and T. Roca-Cortes in Tenerife and E. Rhodes Jr. and colleagues at Mt. Wilson. W.J.C., R.H., Y.E. S.J.H. and E.M. acknowledge the support of the United Kingdom Science and Technology Facilities Council (STFC) through grant ST/V000500/1. The peak-finding computations used in this paper were performed using the University of Birmingham's BlueBEAR HPC service, which provides a High Performance Computing service to the University's research community. See \url{http://www.birmingham.ac.uk/bear} for more details. S.B. acknowledges NASA grant 80NSSC25K7669. This research has made use of NASA's Astrophysics Data System Bibliographic Services.

\section*{Data Availability}

The BiSON time series data analysed here are available at \url{http://bison.ph.bham.ac.uk/opendata}.


\bibliographystyle{mnras.bst}
\bibliography{mncycle1}


\bsp	
\label{lastpage}
\end{document}